# Transformation of kinematical quantities from rotating into static coordinate system

**Dimitar G Stoyanov**

Faculty of Engineering and Pedagogy in Sliven, Technical University of Sofia
59, Bourgasko Shaussee Blvd, 8800 Sliven, BULGARIA

E-mail: dgstoyanov@abv.bg

**Abstract.** In this article the problem for the using of uniformly rotating coordinate system is considered. The correct relations between the kinematical quantities characterizing the motion of a body relative to uniformly rotating system and static coordinate system are obtained.

## 1. Introduction

In this article the problem of rotating coordinate system handling is discussed. It is not a new one, and is available in one form or another in all textbooks of physics at the undergraduate level. The common point of the bigger part of these books is the pointing out of ungrounded interrelations between the kinematical quantities in static and rotating coordinate systems, and these interrelations have been deeply rooted in the consciousness of students and teachers. Nowadays such information is available in Wikipedia [1]. It is essentially that this fact reduces the deepness of rationalization and understanding on behalf of the students of wide circle of basic laws and effects in mechanics.

As an exception is [2], where an attempt is made for the derivation of the kinematical interrelations of coordinate systems disposed in the common case and the kinematical interrelations of the motion of a body also in common case. The approach is true and the final correct result should be got if in a later stage of the derivation a matrix for the transformation of the coordinates has been applied. This is an example how in order the present material to be simplified could be fallen in situation when the present material will be not quite adequate. Only in [3] the correct interrelations are given, moreover combined with a proof in general case.

The objective of this article is following [2] up to a certain place and continuing after that in own way the kinematical interrelations between the quantities in a static and uniformly rotating coordinate systems to be derived correctly.

## 2. Relations between the kinematical quantities
### 2.1 A detail calculation

Let the material point (small body) *M* with mass *m* be moving in space. Let the inertial Cartesian coordinate system *S* to be static. The position of the material point in relation to the coordinate system *S* is determined by the radius-vector $\vec{r}$ (1), where *x, y* and *z* are the coordinates of the geometric point relative to *S* (see figure 1), in which the material point is:

$$\vec{r} = \begin{pmatrix} x \\ y \\ z \end{pmatrix} \qquad (1)$$

Let the Cartesian coordinate system $S'$, shown in figure 1, be rotating. The position of the material point in relation to the coordinate system $S'$ is determined by the radius-vector $\vec{r}'$ (2), where $x'$, $y'$ and $z'$ are the coordinates of the geometric point relative to $S'$, in which the material point is:

$$\vec{r}' = \begin{pmatrix} x' \\ y' \\ z' \end{pmatrix} \quad (2)$$

Suppose the coordinate system $S'$ is such that:
- the origins of the coordinate systems $S'$ and $S$ coincide;
- the axis $z'$ of $S'$ coincides with the axis $z$ of $S$;
- at the initial moment $t = 0$ the axes $OX'$ and $OX$ coincide;
- $S'$ is rotating uniformly around the axis $z$ of $S$ with angular velocity $\Omega$.

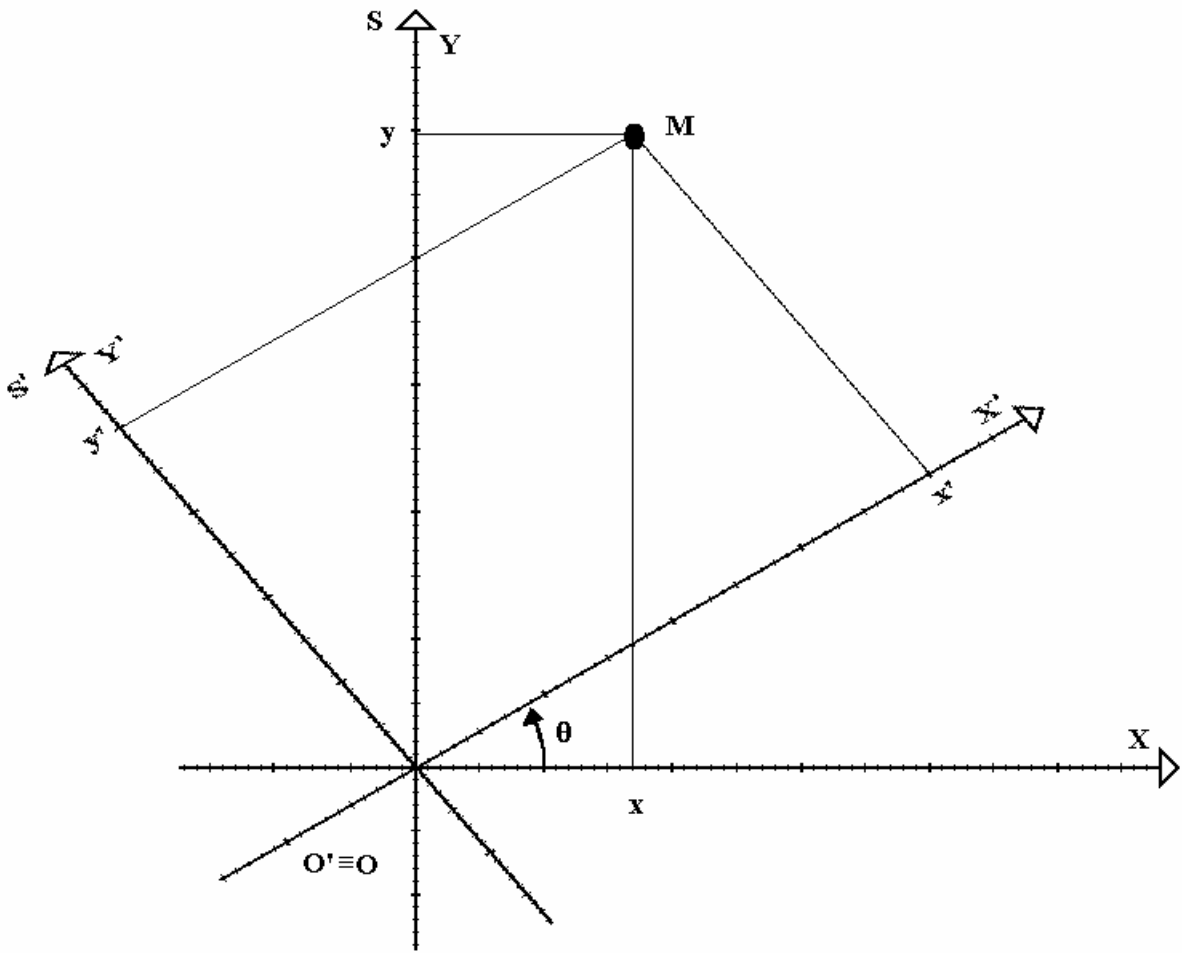

Figure 1. Mutual disposition of the coordinate systems.

A case with simple mutual disposition of the coordinate systems is considered. This allows an intuitive check of the obtained results to be done. The calculations are given deliberately in details in order to preclude any logical misunderstandings. The considered here special case of motion of the rotating coordinate system $S'$ is sufficient the necessary relations between the kinematical quantities to be obtained as an illustration without any loss of generality.

At such formulation the coordinates of the geometric point relative to $S'$ are transformed to the coordinates of the same point relative to $S$ by the equations [2]:

$$x = x'.cos(\Omega.t) - y'.sin(\Omega.t), \qquad (3a)$$
$$y = x'.sin(\Omega.t) + y'.cos(\Omega.t), \qquad (3b)$$
$$z = z'. \qquad (3c)$$

Further we consider the motion of a material point which has radius-vector $\vec{r}'$, velocity $\vec{v}'$ and acceleration $\vec{a}'$ relative to $S'$ and radius-vector $\vec{r}$, velocity $\vec{v}$ and acceleration $\vec{a}$ relative to $S$.

Differentiating (3a), (3b) and (3c) with respect to time we obtain:

$$v_x = v'_x.cos(\Omega.t) - \Omega.x'.sin(\Omega.t) - v'_y.sin(\Omega.t) - \Omega.y'.cos(\Omega.t), \qquad (4a)$$
$$v_y = v'_x.sin(\Omega.t) + \Omega.x'.cos(\Omega.t) + v'_y.cos(\Omega.t) - \Omega.y'.sin(\Omega.t), \qquad (4b)$$
$$v_z = v'_z. \qquad (4c)$$

where $\mathbf{v}_x$, $\mathbf{v}_y$ and $\mathbf{v}_z$ are the components of the velocity $\vec{v}$ of the material point relative to $S$.

Similarly, differentiating (4a), (4b) and (4c) with respect to time for the components of the acceleration $\vec{a}$ of the material point relative to $S$, we obtain:

$$\mathbf{a}_x = a'_x.cos(\Omega.t) - 2.\Omega.v'_x.sin(\Omega.t) - \Omega^2.x'.cos(\Omega.t)$$
$$\qquad - a'_y.sin(\Omega.t) + 2.\Omega.v'_y.cos(\Omega.t) + \Omega^2.y'.sin(\Omega.t), \qquad (5a)$$

$$\mathbf{a}_y = a'_x.sin(\Omega.t) + 2.\Omega.v'_x.cos(\Omega.t) - \Omega^2.x'.sin(\Omega.t)$$
$$\qquad + a'_y.cos(\Omega.t) - 2.\Omega.v'_y.sin(\Omega.t) - \Omega^2.y'.cos(\Omega.t), \qquad (5b)$$

$$\mathbf{a}_z = a'_z. \qquad (5c)$$

Up to here all obtained results coincide with the results presented in [2].

Further on in [2] "a projection" of the quantities from the right sides (concerning the rotating coordinate system $S'$) on the axes of the static coordinate system $S$ is stated.

To make the projection process more clear, we will use the matrix for transformation of coordinates.

## 2.2 A presentation by the transformation matrix

In this part of the article the obtained yet results for the velocity and the acceleration of the material point will be presented by the matrix for transformation of coordinates.

We define a matrix (3x3) for transformation of coordinates $\hat{R}$, derived from (3a), (3b) and (3c):

$$\hat{R} = \begin{pmatrix} cos(\theta) & -sin(\theta) & 0 \\ sin(\theta) & cos(\theta) & 0 \\ 0 & 0 & 1 \end{pmatrix}. \qquad (6)$$

Here $\theta$ is the angle between the axes $OX$ and $OX'$ of the coordinate systems $S$ and $S'$. The positive direction of the angle is counterclockwise (see figure1).

The matrix $\hat{R}$ acts upon vectors represented in $S'$ and as a result gives vectors from $S$.

Because of that

$$\det \hat{R} = 1. \quad (7)$$

the magnitudes of vectors at transformation are not scaled (the length of the line segment is kept during the transition from one coordinate system to another).

If the angle $\theta$ is a constant and is equal to zero $\hat{R}$ is the identity matrix.

If the angle $\theta$ is a constant and is not equal to zero $\hat{R}$ is not the identity matrix. In this case $S'$ is turned relative to $S$, but is static.

If $\theta = \Omega.t$ the system $S'$ rotates with respect to $S$ around axis $z$ with a constant angular velocity $\Omega$. In this case $\hat{R}$ becomes

$$\hat{R} = \begin{pmatrix} \cos(\Omega.t) & -\sin(\Omega.t) & 0 \\ \sin(\Omega.t) & \cos(\Omega.t) & 0 \\ 0 & 0 & 1 \end{pmatrix}. \quad (8)$$

By using of (8) the components of the radius-vector (3a, 3b and 3c), the velocity (4a, 4b and 4c) and the acceleration (5a, 5b and 5c) can be presented in the following form

$$\begin{pmatrix} x \\ y \\ z \end{pmatrix} = \hat{R}. \begin{pmatrix} x' \\ y' \\ z' \end{pmatrix} \quad (9)$$

$$\begin{pmatrix} v_x \\ v_y \\ v_z \end{pmatrix} = \hat{R}. \begin{pmatrix} v'_x - \Omega.y' \\ v'_y + \Omega.x' \\ v'_z \end{pmatrix} \quad (10)$$

$$\begin{pmatrix} a_x \\ a_y \\ a_z \end{pmatrix} = \hat{R}. \begin{pmatrix} a'_x - 2.\Omega.v'_y - \Omega^2.x' \\ a'_y + 2.\Omega.v'_x - \Omega^2.y' \\ a'_z \end{pmatrix} \quad (11)$$

It easy to see, that the right side of the obtained equations (9), (10) and (11) are the products of the matrix $\hat{R}$ by the column vectors of quantities given with respect to the system $S'$. These equations can be written in a compact form using the properties of the vector product of vectors. For the goal a vector of the angular velocity $\vec{\Omega}$ of the coordinate system $S'$ relative to $S$ is defined as in this considerable special case the systems are of the following form

$$\vec{\Omega} = \begin{pmatrix} 0 \\ 0 \\ \Omega \end{pmatrix} \quad (12)$$

Then from (12) we derive:

$$\vec{r} = \hat{R}.\vec{r}' \quad (13)$$

$$\vec{v} = \hat{R}.(\vec{v}' + \vec{\Omega} \times \vec{r}') \tag{14}$$

$$\vec{a} = \hat{R}.[\vec{a}' + 2.\vec{\Omega} \times \vec{v}' + \vec{\Omega} \times (\vec{\Omega} \times \vec{r}')] \tag{15}$$

This is the final form of the kinematical interrelations between the static and the uniformly rotating coordinate system.

### 2.3 Generalization

It is evident from the results obtained above the importance of the matrix for transformation of coordinates $\hat{R}$. The obtained up to now results can illustrate the generalization proved in [3]. This generalization includes the following:

Let's have vector $\vec{A} \in S$ and vector $\vec{A}' \in S'$ for which is valid the following link

$$\vec{A} = \hat{R}.\vec{A}' \tag{16}$$

The matrix $\hat{R}$ is anti-symmetrical that allows the introduction of a vector of the instant angular velocity $\vec{\Omega}$ of rotation of $S'$ relative to $S$ ($\vec{\Omega}$ to be constant is not compulsory), which satisfies the condition [3]

$$\frac{d\hat{R}}{dt}.\vec{A}' = \hat{R}.\hat{R}^{-1}.\frac{d\hat{R}}{dt}.\vec{A}' = \hat{R}.\vec{\Omega} \times \vec{A}' \tag{17}$$

Differentiating with respect to time (16) and using (17) we obtain [3]

$$\frac{d\vec{A}}{dt} = \hat{R}.(\frac{d\vec{A}'}{dt} + \vec{\Omega} \times \vec{A}') \tag{18}$$

The rule (18) is valid for each rotating motion of the coordinate system $S'$ relative to $S$, i.e. with an arbitrary orientation in space, uniformly or non-uniformly in time.

Differentiating (18) with respect to time and making certain transformations and using (17) we can obtain [3]

$$\frac{d^2\vec{A}}{dt^2} = \hat{R}.\left[\frac{d^2\vec{A}'}{dt^2} + \frac{d\vec{\Omega}}{dt} \times \vec{A}' + 2.\vec{\Omega} \times \frac{d\vec{A}'}{dt} + \vec{\Omega} \times (\vec{\Omega} \times \vec{A}')\right]. \tag{19}$$

The comparison between (18) and (14) on one hand, and between (19) and (15) on the other hand shows the accuracy of the obtained previously and considered in details interrelations. This is an illustration of the importance of the transformation matrix $\hat{R}$ as well as the possibility for generalizations if this approach is used.

### 3. Discussion of results

The transformation matrix (6) is an important element at the kinematical interrelations (13), (14) and (15) because it is responsible for the correct direction of vectors in $S$. Unfortunately, its significance is not appreciated properly up to now.

According to the mathematics we can sum, subtract and multiply vectors, which are from one and the same vector space or are presented in one and the same coordinate system. This is the reason in [3] several times to be reminded that $\vec{r}' \in S'$ and $\hat{R}.\vec{r}' \in S$ are two different vectors belonging to two different vector spaces.

From (14) is easily obtained that

$$\vec{v} = \hat{R}.\vec{v}' + \vec{\Omega} \times \vec{r} \neq \vec{v}' + \vec{\Omega} \times \vec{r} \qquad (20)$$

The transformation of $\vec{v}'$ in (20) by $\hat{R}$ is necessary not only for the uniformly rotating coordinate system but for the turned system $S'$.

If we multiply both sides of (15) by the mass of the material point and taking into account that according to the second law of Newton

$$m.\vec{a} = \vec{F} = \hat{R}.\vec{F}' \qquad (21)$$

where $\vec{F}$ is the force acting on the material point we will obtain:

$$m.\vec{a}' = \vec{F}' - 2.m.\vec{\Omega} \times \vec{v}' - m.\vec{\Omega} \times (\vec{\Omega} \times \vec{r}') \qquad (22)$$

Thus, in the right side of (22) the Coriolis and centrifugal inertial forces. It is acting on the body from the point of view of uniformly rotating coordinate system $S'$, have appeared naturally and correctly. From (22) is evident that if there is no force $\vec{F}$, the material point will move relative to $S'$ with acceleration which value differs from zero. The reason is the presence of Coriolis and centrifugal inertial forces. Exactly these inertial forces make the uniformly rotating coordinate system $S'$ non-inertial.

**4. Conclusion**
In this article the geometric interrelations are precised and the common case of motion of a body is considered. The conclusion is elementary and easy of access for students and therefore, it can be used in the teaching activity.

The considered approach is valuable not only with that it gives the right form of the kinematical interrelations but further developed naturally can give the expressions for the interrelations between the dynamical quantities for a single material point, a system of material points and a rigid body in various coordinate systems